\documentclass[12pt]{article}
\usepackage[margin=1in]{geometry}
\usepackage{amsmath}
\usepackage{amsthm}
\usepackage{amssymb}
\usepackage{bm}
\usepackage[]{hyperref}
\usepackage{graphicx}

\numberwithin{equation}{section}

\def\mysavedown#1{\edef\mysubs{\mysubs#1}}
\def\mysaveup#1{\edef\mysups{\mysups#1}}
\def\mydown#1{{\mytensor}_{\vphantom{\mysubs}#1}}
\def\myup#1{{\mytensor}^{\vphantom{\mysups}#1}}
\def\tensor#1#2{
  #1
  \def\mytensor{\vphantom{#1}}
  \def\mysubs{\relax}
  \def\mysups{\relax}
  \let\down=\mysavedown
  \let\up=\mysaveup
  #2
  \let\down=\mydown
  \let\up=\myup
  #2
  }

\newcommand{\xxsigma}{d\sigma^{1}}
\newcommand{\xxtau}{d\sigma^{0}}

\DeclareMathOperator{\rank}{rank}

\newcommand{\hodge}{\ast}
\newcommand{\Atil}{\widetilde{A}}
\newcommand{\Btil}{\widetilde{B}}
\newcommand{\Ctil}{\widetilde{C}}

\newcommand{\Ftil}{\widetilde{F}}
\newcommand{\Htil}{\widetilde{H}}
\newcommand{\Mtil}{\widetilde{M}}

\newcommand{\Qtil}{\widetilde{Q}}

\newcommand{\Util}{\widetilde{U}}
\newcommand{\Vtil}{\widetilde{V}}

\newcommand{\bbR}{\mathbb{R}}

\newcommand{\htil}{\tilde{h}}
\newcommand{\half}{\frac{1}{2}}

\newcommand{\lieg}{\mathfrak{g}}

\newcommand{\ntil}{\tilde{n}}

\newcommand{\omegatil}{\tilde{\omega}}
\newcommand{\pitil}{\tilde{\pi}}

\newcommand{\thetatil}{\tilde{\theta}}

\newcommand{\util}{\tilde{u}}
\newcommand{\vtil}{\tilde{v}}

\newcommand{\xstil}[1]{\xtil^{#1}{}_{\sigma}}
\newcommand{\xs}[1]{x^{#1}{}_{\sigma}}
\newcommand{\xttil}[1]{\xtil^{#1}{}_{\tau}}
\newcommand{\xt}[1]{x^{#1}{}_{\tau}}
\newcommand{\xtil}{\tilde{x}}
\newcommand{\xitil}{\tilde{\xi}}

\begin{document}
%title
\begin{titlepage}
\noindent
\strut\mbox{December 2007}
\par\vskip 2cm
\begin{center}
    {\Large \bf Target Space Duality III: Potentials}\\[0.5in]
    %author
    {\bf Orlando Alvarez}\footnote{email: \tt oalvarez@miami.edu}\\
    {\bf Blazej Ruszczycki}\footnote{email: \tt ruszczycki@physics.miami.edu}\\[0.1in]
    %address
    Department of Physics\\
    University of Miami\\
    P.O. Box 248046\\
    Coral Gables, FL 33124 USA\\[0.3in]
\end{center}
\par\strut\vspace{.5in}
\noindent

%abstract
\begin{abstract}
    We generalize previous results on target space duality to the case
    where there are background fields and the sigma model lagrangian
    has a potential function.
\end{abstract}

\vspace{.25in}

\end{titlepage}

% Begin main text
\section{Introduction}
\label{sec:introduction}

In two articles \cite{Alvarez:2000bh,Alvarez:2000bi} henceforth
referred to as Paper~I and Paper~II respectively, a general theory was
developed for irreducible target space duality\footnote{For a list of
references see Papers I and II.} for classical sigma models
characterized by a target space $M$, a riemannian metric $g$ and an
antisymmetric tensor field $B$.  By irreducible duality we mean that
all fields participate in the duality transformation and that there
are no spectator fields.  This rules out the derivation of the
important Buscher formulas \cite{Buscher:1988qj} and also duality in
WZW models \cite{Witten:1984ar} where the duality transformation is
performed by gauging an anomaly free subgroup, e.g., see the
discussions in
\cite{Kiritsis:1991zt,Rocek:1992ps,Gasperini:1993nz,delaOssa:1993vc,%
Giveon:1994ai,Alvarez:1994qi,Alvarez:1994zr,Sfetsos:1994vz}.  The
latter remark requires some explanation.  For example, consider a WZW
model on a compact simple Lie group $G$ where the diagonal subgroup
$G_{D}$ of the symmetry group $G\times G$ is the anomaly free subgroup
that is gauged.  Schematically, the prescription to construct the dual
model is that the original model with fields $g$ is augmented to an
equivalent $G_{D}$ gauge invariant model with fields $g,A,\lambda$
where $A$ are $G_{D}$ gauge fields and $\lambda$ are Lagrange
multipliers that enforce the vanishing of the field strengths.  In
principle the idea is to eliminate the variables $g$ and $A$ in favor
of the Lagrange multipliers $\lambda$.  Naively the original model
with variables $g$ had $\dim G$ degrees of freedom.  The dual model
with variables $\lambda$ would also have $\dim G$ degrees of freedom. 
Unfortunately this procedure does not work for a variety of reasons. 
If  by brute force we attempt to eliminate the variables $g,A$ then
the action for the $\lambda$ fields is nonlocal.  We can try to
finesse things by using the gauge invariance of the theory $g \to
hgh^{-1}$.  Unfortunately this does not allow us to gauge $g$ to the
identity element.  The best we can do is gauge $g$ to an element $t$
of a maximal torus $T$ and we have a residual $T$ gauge invariance. 
This residual gauge invariance can be used to gauge $\rank G$ of the
Lagrange multipliers to zero \cite{delaOssa:1993vc}.  The $A$
variables can be eliminated and we are left with a local action
involving only $t$ and the remaining Lagrange multipliers.  We note
that the ``$t$'' variables are spectators and thus the methods of
Papers I and II do not apply.  See the worked out example in
\cite{Giveon:1994ai}.  Finally we mention why the results of Papers I
and II suggest that it is impossible to eliminate the variables $g,A$
in favor of a local action involving only the Lagrange multipliers
$\lambda$.  If $\lieg$ denotes the Lie algebra of $G$ then the
Lagrange multipliers take values in $\lieg^{*}$, the dual vector
space.  This strongly suggests that the duality transformation here is
related to the cotangent bundle $T^{*}G = G \times \lieg^{*}$.  In
Papers I and II we showed that duality associated with any cotangent
bundle $T^{*}M$ implied that $M$ is a compact Lie group with the
$3$-form $H$ being exact in a very specific way.  The $3$-form in the
WZW model is topologically nontrivial.

\section{Framework}
\label{sec:framework}

The sigma model with target space $M$, metric $g$, $2$-form $B$ and
potential function $U$ will be denoted by $(M,g,B,U)$ and has
lagrangian density
\begin{align}
    \mathcal{L} & = \half g_{ij}(x)
    \left(
    \frac{\partial x^{i}}{\partial \tau}
    \frac{\partial x^{j}}{\partial \tau}
    -\frac{\partial x^{i}}{\partial \sigma}
    \frac{\partial x^{j}}{\partial \sigma}
    \right) +
    B_{ij}(x)
    \frac{\partial x^{i}}{\partial \tau}
    \frac{\partial x^{j}}{\partial \sigma}
    - U(x)
    \nonumber \\ 
    & + 
    A_{i}(x) \frac{\partial x^{i}}{\partial \tau}
    + C_{i}(x) \frac{\partial x^{i}}{\partial \sigma}.
    \label{eq:lag} 
\end{align}
We will generally follow the notation and formalism introduced in
\cite{Alvarez:2000bh,Alvarez:2000bi}.  In the above we have introduced
two background fields $A_{i}$ and $C_{i}$ that break worldsheet
Lorentz invariance for the following reason.  Assume we have a Lorentz
invariant field theory with fields $(x^{i},y^{a})$ but where we are
not interested in irreducible duality.  Namely, only the fields
$x^{i}$ participate in the duality transformation and the $y^{a}$
fields are spectators.  In this case we can regard the fields $y^{a}$
as parameters and $g$, $B$ and $U$ depend on the $y^{a}$
parametrically.  In the full lagrangian density there may be a term of
type $K_{ai}(x,y)\partial y^{a} \partial x^{i}$ and this will become a
contribution to the second line of \eqref{eq:lag} when the fields
$y^{a}$ are held fixed.  The canonical momentum density is
\begin{equation}
    \pi_{i}= \frac{\partial\mathcal{L}}{\partial \dot{x}^{i}} =
    g_{ij}\dot{x}^{j} + B_{ij}x'{}^{j} + A_{i}(x)\;.
    \label{eq:canmomentum}
\end{equation}
The hamiltonian density may be written as 
\begin{align}
    \mathcal{H} & =  \half g^{ij}
    \left(\pi_{i} - B_{ik}\frac{dx^{k}}{d\sigma}\right)
     \left(\pi_{j} - B_{jl}\frac{dx^{l}}{d\sigma}\right)
   + \half g_{ij} \frac{dx^{i}}{d\sigma}\frac{dx^{j}}{d\sigma}
   \nonumber \\
   &+ U(x)  + \half g^{ij} A_{i}A_{j}
    -  g^{ij}A_{i}\left(\pi_{j} - B_{jl}\frac{dx^{l}}{d\sigma}\right)
   - C_{i}\frac{dx^{i}}{d\sigma}\;.
    \label{eq:hamiltonian}
\end{align}

We are interested in studying duality between sigma models with
lagrangian densities of the type \eqref{eq:lag}.  Here we consider a
generalization of the canonical transformations considered in
Paper~I that still leads to a linear relationship
between the respective $\pi$ and $dx/d\sigma$ in the two models.  The
generating function $F$ for the duality canonical transformation will
be of the form
\begin{equation}
    F[x(\sigma),\xtil(\sigma)] = \int \alpha +
    \int u(x,\xtil)d\sigma\,,
    \label{eq:genfunction}
\end{equation}
where $\alpha$ is a $1$-form on $M\times\Mtil$, see
Paper~I, and $u$ a function on $M\times\Mtil$.  The
canonical transformation is given by
\begin{align*}
    (\pi - B x')_{i} & =  m_{ji}\frac{d\xtil^{j}}{d\sigma} +
    n_{ij}\frac{dx^{j}}{d\sigma} - \frac{\partial u}{\partial x^{i}}\,,
    \\
    (\pitil - \Btil \xtil')_{i} & =  m_{ij}\frac{dx^{j}}{d\sigma} +
    \ntil_{ij}\frac{d\xtil^{j}}{d\sigma} + \frac{\partial u}{\partial
    \xtil^{i}}\,,
\end{align*}
where $m_{ij}$ and $n_{kl}$ will be discussed shortly.

It is best to now go to orthonormal coframes on $M$ and $\Mtil$.  Let
$(\theta^{1},\ldots,\theta^{n})$ be a local orthonormal
coframe\footnote{Because we will be working in orthonormal frames we
do not distinguish an upper index from a lower index in a tensor.} for
$M$.  The Cartan structural equations are
\begin{align*}
    d\theta^{i} & =  -\omega_{ij}\wedge\theta^{j}\;, \\
    d\omega_{ij} & =  -\omega_{ik}\wedge\omega_{kj}
    +\half R_{ijkl}\theta^{k}\wedge\theta^{l}\;,
\end{align*}
where $\omega_{ij}=-\omega_{ji}$ is the unique torsion free riemannian
connection associated with the metric $g$.  We remind the reader that
the analog of $dx^{i}/d\sigma$ in an orthonormal coframe is $\xs{i}$
defined by $\theta^{i}=\xs{i}d\sigma$.  There are similar definitions
pertaining to $\Mtil$.

Following the discussion in Paper~I, the canonical
transformation may be expressed in terms of a $2$-form $\gamma$
closely related to $d\alpha$ on $M\times\Mtil$ and given by
\begin{equation}
    \gamma = -\half n_{ij}(x,\xtil) \theta^{i}\wedge \theta^{j} + 
    m_{ij}(x,\xtil) \thetatil^{i}\wedge \theta^{j} + \half 
    \ntil_{ij}(x,\xtil) \thetatil^{i}\wedge \thetatil^{j} \,.   
    \label{eq:gammaframe}
\end{equation}
The $2$-form $\gamma$ is not closed but satisfies
\begin{equation}
    d\gamma = H - \Htil
    \label{eq:dgamma}
\end{equation}
where $H=dB$ and $\Htil=d\Btil$.  The derivatives of the function $u$
are given in the orthonormal frame by\footnote{\label{foot:sign} Note
the unconventional negative sign in the definition above.  This is
introduced to make subsequent equations more symmetric.}
\begin{equation}
    du = u_{i} \theta^{i} - \util_{i}\thetatil^{i}\,.
    \label{eq:defdu}
\end{equation}
In terms of the orthonormal frame the canonical transformation may be
written as
\begin{align}
    (\pi - B \xs{\relax})_{i} & =  m_{ji}\xstil{j} +
    n_{ij}\xs{j} - u_{i}\,, 
    \label{eq:pitransf} \\
    (\pitil - \Btil \xstil{\relax})_{i} & =  m_{ij}\xs{j} +
    \ntil_{ij}\xstil{j} - \util_{i}\,,
    \label{eq:pitiltransf}
\end{align}
where we now interpret the components of $\pi,\pitil,
B,\Btil,m,n,\ntil$ to be given with respect to the orthonormal
frames.

We digress for a second and make a general observation.  Assume we
have equal dimensional vector spaces $V$ and $\Vtil$ where we use the
notation $(\cdot,\cdot)$ for the inner product on either space.  Let
$L:V\to\Vtil$ be an invertible linear transformation and let $Q(v) =
\half (v,v) + (a,v) + b$ be a real value quadratic function on $V$. 
There is a corresponding quadratic function $\Qtil$ on $\Vtil$. 
Assume we are told that the affine transformation $\vtil = Lv +
\tilde{w}$ maps $Q$ into $\Qtil$.  A short computation shows that
$$
    \half (Lv,Lv) + (\tilde{a} +\tilde{w},Lv) + \tilde{b} +
    (\tilde{a},\tilde{w}) + \half (\tilde{w},\tilde{w}) = \half (v,v)
    + (a,v) + b\,.
$$
Comparing both sides we would conclude that $L$ is an isometry,
$\tilde{a} +\tilde{w} = La$ and $b = \tilde{b} + (\tilde{a},\tilde{w})
+ \half (\tilde{w},\tilde{w})$.

In our case we require the canonical transformation to preserve the 
hamiltonian densities up to a total derivative
\begin{equation}
    \widetilde{\mathcal{H}} = \mathcal{H} + \frac{dh}{d\sigma}\;,
    \label{eq:Htransf}
\end{equation}
where $h$ is a function on $M\times\Mtil$.  Our canonical
transformation, given by \eqref{eq:pitransf} and 
\eqref{eq:pitiltransf}, is an  affine  mapping of $(\xs{\relax},\pi)$ into
$(\xstil{\relax},\pitil)$. We can use parts of the general argument
about the transformation of quadratic functions given in the previous
paragraph.  We have to be careful because the derivative in
\eqref{eq:Htransf} modifies some of the conclusions of the previous
paragraph.  The linear part of the transformation must be an isometry,
a restriction studied in Paper~I, where we found
\begin{align}
    mm^{t} & =  I - \ntil^{2}\;,
    \label{eq:mmt}  \\
    m^{t}m & =  I- n^{2}\;,
    \label{eq:mtm}  \\
    -mn & =  \ntil m\;.
    \label{eq:mn}
\end{align}
If we write\footnote{See footnote \ref{foot:sign}.} $dh =
h_{i}\theta^{i} - \htil_{i}\thetatil^{i}$ then $dh/d\sigma =
h_{i}\xs{i} - \htil_{i}\xstil{i}$.  From this we learn that
\begin{align*}
    h_{i} & =  C_{i} - m_{ji}(\util_{j} + \Atil_{j}) -
    n_{ij}(u_{j}+A_{j})\,,
    \\
    \htil_{i} & =  \Ctil_{i} -m_{ij}(u_{j}+A_{j}) -
    \ntil_{ij}(\util_{j} +\Atil_{j})\,.
\end{align*}

The problem we have to solve is to find functions $u,h: M\times\Mtil 
\to \bbR$ such that  
\begin{align}
    du & =  u_{i}\theta^{i} - \util_{i}\thetatil^{i}\,,
    \label{eq:duresult} \\
    dh & =  \left(
    C_{i} - m_{ji}(\util_{j} + \Atil_{j}) -
    n_{ij}(u_{j}+A_{j}) \right) \theta^{i}
    \nonumber  \\
     &\quad -  \left( \Ctil_{i} -m_{ij}(u_{j}+A_{j}) -
    \ntil_{ij}(\util_{j} +\Atil_{j}) \right) \thetatil^{i}\,.
    \label{eq:dhresult}
\end{align}
The integrability equations for the system given above, $d^{2}u=0$ and
$d^{2}h=0$, lead to hyperbolic PDEs for $u$.  Finally we observe that
there is one more relation that must be satisfied for
\eqref{eq:Htransf} to hold:
\begin{equation}
    \Util + \half (\util_{j} + \Atil_{j})(\util_{j} + \Atil_{j})
    = U + \half (u_{j}+A_{j})(u_{j}+A_{j})\,.
    \label{eq:UUtil}
\end{equation}
Remember that $U$ is a function on $M$ and $\Util$ is a functions on
$\Mtil$ so $dU= U_{i}\theta^{i}$ and $d\Util =
\Util_{i}\thetatil^{i}$.

The generating function for canonical transformations is only locally
defined.  We could ask whether it is possible to give a more global
formulation.  We think this is possible.  Notice that of primary
interest to us is not the function $u$ but its derivatives.  For this
reason it is convenient to ``define''
\begin{equation}
    v_{i} = u_{i}+ A_{i}\quad\mbox{and}\quad \vtil_{j}=\util_{j} +
    \Atil_{j}\,.
    \label{eq:defvvtil}
\end{equation}
Let $F_{A}$ be the curvature $A = A_{i}\theta^{i}$ and $\Ftil_{\Atil}$
be the curvature of $\Atil= \Atil_{i}\thetatil^{i}$.  Consider a
$1$-form on $M\times\Mtil$ defined by
\begin{equation}
    \xi = v_{i}\theta^{i} - \vtil_{i}\thetatil^{i}\,.
    \label{eq:defxi}
\end{equation}
The equation $d^{2}u=0$ is replaced by
\begin{equation}
    d\xi = F_{A} - \Ftil_{\Atil}\,.
    \label{eq:dxi}
\end{equation}
In a similar fashion the equation $d^{2}h=0$ is replaced by
\begin{equation}
    d\left[
      -(m_{ji}\vtil_{j} + n_{ij}v_{j}) \theta^{i}
      +( m_{ij}v_{j} + \ntil_{ij}\vtil_{j})\thetatil^{i}
    \right] = -F_{C} + \Ftil_{\Ctil}\,,
    \label{eq:dxidual}
\end{equation}
where $F_{C}$ is the curvature of $C = C_{i}\theta^{i}$ and
$\Ftil_{\Ctil}$ is the curvature of $\Ctil = \Ctil_{i}\thetatil^{i}$. 
Similarly the equation for the potentials becomes
\begin{equation}
    \Util + \half \vtil_{i}\vtil^{i} =
    U + \half v_{i}v^{i}\,.
    \label{eq:newU}
\end{equation}

\section{Pseudoduality}
\label{sec:pseudoduality}
Here we switch to the framework where we consider the map between the paths on one manifold and the paths on the other.
We use directly \eqref{eq:pitransf},\eqref{eq:pitiltransf} and \eqref{eq:mmt} to \eqref{eq:mn}, 
having in mind the rest of the discussion as a guideline.
In this way we work directly with equations of motion what makes the calculations more straightforward; we have a system
of PDE's for which we obtain the integrability conditions. 
Moreover, for the 2-dimensional space the Hodge duality transforms 1-forms into another 1-forms. We may use it to write the
equation in geometrical, covariant fashion.   
    
We restrict to the case $A=C=0$. Introducing the lightcone coordinates $\sigma^\pm=\tau \pm \sigma$ the equations of motion for
lagrangian \eqref{eq:lag} are
\begin{equation}
    x^{k}_{+-} = - \half\, H_{kij} x^{i}{}_{+} x^{j}{}_{-}
    - \frac{1}{4} \, U_{k}\,,
    \label{eq:eom}
\end{equation}
where $dU = U_{k} \theta^{k}$.

We rewrite the duality transformations  \eqref{eq:pitransf}
and \eqref{eq:pitiltransf} in terms of the velocities as
\begin{align}
    \xt{i} & =  m_{ji}\xstil{j} +
    n_{ij}\xs{j} - u_{i}\,, 
    \label{eq:xdottransf} \\
    \xttil{i} & =  m_{ij}\xs{j} +
    \ntil_{ij}\xstil{j} - \util_{i}\,.
    \label{eq:xtildottransf}
\end{align}
Mimicking the computations of Section~3 of \cite{Alvarez:2000bh} we
find that
\begin{equation}
    \begin{pmatrix}
        m^{t} & 0  \\
        -\ntil & I
    \end{pmatrix}
    \begin{pmatrix}
        \xstil{\relax}  \\
        \xttil{\relax} + \util
    \end{pmatrix}
      = 
     \begin{pmatrix}
         -n & I  \\
         m & 0
     \end{pmatrix}
     \begin{pmatrix}
         \xs{\relax}  \\
         \xt{\relax} + u
     \end{pmatrix} .
    \label{eq:ps1}  
\end{equation}
We can now mimic the discussion in Section~1 of \cite{Alvarez:2000pk}
and restrict ourselves to the special case $T_{+}= T_{-}$. If we lift 
to the frame bundle as discussed in \cite{Alvarez:2002mg} the
pseudoduality equations become
\begin{equation}
    \xtil_{\pm}^{i} + \half \util^{i} =
    \pm\left( x_{\pm}^{i} + \half u^{i} \right)\,.
    \label{eq:psd}
\end{equation}
Using the notation from Section~7 of \cite{Alvarez:2002mg} we have the
equations of motion may be written as
\begin{equation}
    d(\hodge \xi^{i}) + \xi_{ij} \wedge (\hodge \xi^{j})
    = \half h_{ijk} \xi^{j} \wedge \xi^{k} - U_{i}\, d\sigma^{0}
    \wedge d\sigma^{1}\,,
    \label{eq:eom-xi}
\end{equation}
and the duality equations as
\begin{equation}
    \xitil^{i} + \util^{i}\,d\sigma^{0} =
    \hodge \left (\xi^{i} + u^{i} \, d\sigma^{0} \right)\,.
    \label{eq:psd2}
\end{equation}
More explicitly we have that
\begin{align}
    \xitil^{i} &= \hodge \xi^{i}+ u^{i}\,d\sigma^{1} -
    \util^{i}\,d\sigma^{0}\,, 
    \label{eq:psd3}\\
    \xi^{i} &= \hodge \xitil^{i} + \util^{i}\,d\sigma^{1} -
    u^{i}\, d\sigma^{0}\,.
    \label{eq:psd4}
\end{align}

Next we define the covariant derivatives of $u_{i}$ and $\util_{i}$ by
\begin{equation}
    \begin{split}
	du_{i} + \omega_{ij}u^{j} & = u'_{ij}\omega^{j} + u''_{ij}
	\omegatil^{j}\,, \\
	d\util_{i} + \omegatil_{ij}\util^{j} & = \util'_{ij}\omega^{j} +
	\util''_{ij} \omegatil^{j}\,.
	\end{split}
    \label{eq:covd-u}
\end{equation}
From $d^{2}u=0$ we see that
\begin{align}
    u'_{ij} & = u'_{ji}\,,
    \label{eq:up-sym}  \\
    \util''_{ij} & = \util''_{ji}\,,
    \label{eq:utilpp-sym}  \\
    u''_{ij} & = - \util'_{ji}\,.
    \label{eq:u-util}
\end{align}
The reason for the unusual negative sign in \eqref{eq:u-util} is the
unconventional definition \eqref{eq:defdu}.

To determine conditions necessitated for the duality equations we
study the integrability conditions on \eqref{eq:psd3} by taking its
exterior derivative
\begin{equation*}
    \begin{split}
    0 &= -\frac{1}{2}
    \tensor{h}{\up{i}\down{j}\down{k}} \tensor{\xi}{\up{j}}\wedge
    \tensor{\xi}{\up{k}} - \tensor{u'}{\up{i}\down{j}}
     \tensor{\xi}{\up{j}}\wedge \xxsigma +
    \tensor{\tilde{u}'}{\up{i}\down{j}} \tensor{\xi}{\up{j}}\wedge
    \xxtau 
    \\
    & \quad 
    - \tensor{u''}{\up{i}\down{j}}
    \tensor{\tilde{\xi}}{\up{j}}\wedge \xxsigma +
    \tensor{\tilde{u}''}{\up{i}\down{j}}
    \tensor{\tilde{\xi}}{\up{j}}\wedge \xxtau 
    \\
    &\quad -
    \tensor{\tilde{\xi}}{\up{j}}\wedge \tensor{\xi}{\up{i}\down{j}} +
    \tensor{\tilde{\xi}}{\up{j}}\wedge
    \tensor{\tilde{\xi}}{\up{i}\down{j}} 
    \\
    &\quad - \tensor{V}{\up{i}}
    \xxsigma\wedge \xxtau - \tensor{\tilde{u}}{\down{j}} \xxtau\wedge
    \tensor{\xi}{\up{i}\up{j}} + \tensor{\tilde{u}}{\down{j}}
    \xxtau\wedge \tensor{\tilde{\xi}}{\up{i}\up{j}}
    \end{split}
\end{equation*}
As in \cite{Alvarez:2002mg} we identify the orthogonal groups in  the 
two frame bundles by requiring that
\begin{equation}
    \omegatil_{ij} + \half H_{ijk}\omegatil^{k} = \omega_{ij} + \half 
    \Htil_{ijk} \omega^{k}\,.
    \label{eq:conn-id}
\end{equation}
Substituting this into the equation above leads to
\begin{equation*}
    \begin{split}
    0 &=  -\frac{1}{2}
    \tensor{h}{\up{i}\down{j}\down{k}} \tensor{\xi}{\up{j}}\wedge
    \tensor{\xi}{\up{k}} + \frac{1}{2}
    \tensor{\tilde{h}}{\up{i}\down{j}\down{k}}
    \tensor{\xi}{\up{j}}\wedge \tensor{\tilde{\xi}}{\up{k}} 
    \\
    &\quad 
    -
    \tensor{u'}{\up{i}\down{j}} \tensor{\xi}{\up{j}}\wedge \xxsigma +
    \tensor{\tilde{u}'}{\up{i}\down{j}} \tensor{\xi}{\up{j}}\wedge
    \xxtau 
    \\
    &\quad
    - \frac{1}{2} \tensor{\tilde{h}}{\up{i}\down{j}\down{k}}
    \tensor{\tilde{u}}{\up{j}} \tensor{\xi}{\up{k}}\wedge \xxtau -
    \frac{1}{2} \tensor{h}{\up{i}\down{j}\down{k}}
    \tensor{\tilde{\xi}}{\up{j}}\wedge \tensor{\tilde{\xi}}{\up{k}} -
    \tensor{u''}{\up{i}\down{j}} \tensor{\tilde{\xi}}{\up{j}}\wedge
    \xxsigma 
    \\
    &\quad
    + \tensor{\tilde{u}''}{\up{i}\down{j}}
    \tensor{\tilde{\xi}}{\up{j}}\wedge \xxtau + \frac{1}{2}
    \tensor{h}{\up{i}\down{j}\down{k}} \tensor{\tilde{u}}{\up{j}}
    \tensor{\tilde{\xi}}{\up{k}}\wedge \xxtau - \tensor{V}{\up{i}}
    \xxsigma\wedge \xxtau
    \end{split}
\end{equation*}
Next we use the following Hodge duality relations
\begin{equation*}
    \begin{split}
    \xi^{j}\wedge \xi^{k} & = 
    -(\tensor{\hodge{\xi}}{\up{j}}\wedge
    \tensor{\hodge{\xi}}{\up{k}})\,,
    \\
    (\hodge\tilde{\xi}^{i}) \wedge \xxsigma &= 
    -\tensor{\tilde{\xi}}{\up{i}}\wedge \xxtau\,,
    \\
    (\hodge\tilde{\xi}^{i}) \wedge \xxtau &= 
    -\tensor{\tilde{\xi}}{\up{i}}\wedge \xxsigma\,,
    \\
     (\hodge\tilde{\xi}^{j}) \wedge \tilde{\xi}^{k} &= 
     (\hodge\tilde{\xi}^{k}) \wedge \tilde{\xi}^{j}\,,
     \end{split}
\end{equation*}
that we substitute into the integrability conditions to obtain
\begin{equation*}
    \begin{split}
    0 &= -\tensor{u''}{\up{i}\down{j}}
    \tensor{\tilde{\xi}}{\up{j}}\wedge \xxsigma -
    \tensor{\tilde{u}'}{\up{i}\down{j}}
    \tensor{\tilde{\xi}}{\up{j}}\wedge \xxsigma 
    +
    \tensor{h}{\up{i}\down{j}\down{k}} \tensor{u}{\up{j}}
    \tensor{\tilde{\xi}}{\up{k}}\wedge \xxsigma 
    \\
    & \quad
    +
    \tensor{u'}{\up{i}\down{j}} \tensor{\tilde{\xi}}{\up{j}}\wedge
    \xxtau + \tensor{\tilde{u}''}{\up{i}\down{j}}
    \tensor{\tilde{\xi}}{\up{j}}\wedge \xxtau 
    + \frac{1}{2}
    \tensor{\tilde{h}}{\up{i}\down{j}\down{k}} \tensor{u}{\up{j}}
    \tensor{\tilde{\xi}}{\up{k}}\wedge \xxtau - \frac{1}{2}
    \tensor{h}{\up{i}\down{j}\down{k}} \tensor{\tilde{u}}{\up{j}}
    \tensor{\tilde{\xi}}{\up{k}}\wedge \xxtau 
    \\
    & \quad
    - \tensor{u}{\down{j}}
    \tensor{u'}{\up{i}\up{j}} \xxsigma\wedge \xxtau -
    \tensor{h}{\up{i}\down{j}\down{k}} \tensor{u}{\up{j}}
    \tensor{\tilde{u}}{\up{k}} \xxsigma\wedge \xxtau +
    \tensor{\tilde{u}}{\down{j}} \tensor{\tilde{u}'}{\up{i}\up{j}}
    \xxsigma\wedge \xxtau - \tensor{V}{\up{i}} \xxsigma\wedge \xxtau
    \,.
    \end{split}
\end{equation*}
Extracting the information contained in the equation above and the
ones
obtained by taking the exterior derivative of \eqref{eq:psd4} we see
that
\begin{align}
    u''_{ij} + \util'_{ij} + h_{ijk}u^{k} & =0 \,,
    \label{eq:i1}  \\
    u'_{ij} + \util''_{ij} - \half \htil_{ijk}u^{k} + \half h_{ijk}
    \util^{k} & =0\,,
    \label{eq:i2}  \\
    U_{i} + u'_{ij}u^{j} - \util'_{ij}\util^{j} + h_{ijk} u^{j}
    \util^{k} & = 0\,,
    \label{eq:i3}  \\
    \util'_{ij} + u''_{ij} + \htil_{ijk} \util^{k} & = 0 \,,
    \label{eq:i4} \\
    \util''_{ij} + u'_{ij} - \half h_{ijk} \util^{k} + \half
    \htil_{ijk}u^{k} & = 0\,,
    \label{eq:i5} \\
    \Util_{i} + \util''_{ij} \util^{j} - u''_{ij}u^{j} + \htil_{ijk}
    \util^{j} u^{k} & = 0 \,.
    \label{eq:i6}
\end{align}
From the above we immediately learn that
\begin{align}
    h_{ijk}u^{k} & = \htil_{ijk}\util^{k}\,,
    \label{eq:j1}  \\
    h_{ijk}\util^{k} & = \htil_{ijk}u^{k}\,,
    \label{eq:j2}  \\
    h_{ijk}u^{j}\util^{k} & =0\,.
    \label{eq:j3}
\end{align}
An important observation that follows from the above is that if we go 
back to $M \times \Mtil$ then we expect that $u$ should be a function 
of both sets of variables, \emph{i.e.} a nontrivial
function on $M\times \Mtil$.

\begin{align}
    u'_{ij}-u'_{ji} & =0\,,
    \label{eq:s1}  \\
    u''_{ij}-u''_{ji} + h_{ijk}u^{k} & =0\,,
    \label{eq:s5}  \\
    u'_{ij} + \util''_{ij} & =0\,,
    \label{eq:s4}  \\
    u''_{ij} + \util'_{ji} & =0\,,
    \label{eq:s3}  \\
    \util''_{ij}-\util''_{ji} & =0\,,
    \label{eq:s2}  \\
    U_{i} + u'_{ij}u^{j} + u''_{ij}\util^{j} & =0\,,
    \label{eq:s6}  \\
    \Util_{i} + \util'_{ij}u^{j} + \util''_{ij}\util^{j} & =0\,.
    \label{eq:s7}
\end{align}

The above is consistent with the condition that on $M \times \Mtil$ 
we require that
\begin{equation}
    \left(U + \half u_{i}u^{i} \right) = \left(\Util + \half
    \util_{i}\util^{i} \right) + \text{constant},
    \label{eq:s8}
\end{equation}
in agreement with \eqref{eq:UUtil}
 %\textcolor{blue}{This equation is stronger than Blazej's result because it says that $U + \half(u^{2}- \util^{2})$ is precisely $\Util$.}

\section{Conclusions}
We obtained a set of nonlinear algebraic equations in a sense they do not contain the derivatives of 
$x^i$ or $\tilde{x}^i$. The geometric condition \eqref{eq:conn-id} on connections on $M$ and $\Mtil$  
is unchanged by the presence of the potentials and the conclusions from \cite{Alvarez:2002mg} hold in the case discussed here.
In addition we have equations \eqref{eq:s1} to \eqref{eq:s7} involving second derivatives of the generating function $u$ and 
the derivatives of the potential. Using  \eqref{eq:s1} to \eqref{eq:s2} we can integrate \eqref{eq:s7} and \eqref{eq:s8} 
obtaining \eqref{eq:s8} which has already appeared as a condition \eqref{eq:newU} for hamiltonian density to be preserved . 
This condition has not been used in the following derivation. Having the solution of equations of motion on $M$
the condition \eqref{eq:s8} is a constraint for the solution on 
$\Mtil$. There could be however a choice of generating function $u$ by which the constraint is satisfied automatically. 
If with an appropriate choice of coordinate system we have $r$ coordinates for which $u_i=\partial u / \partial x^i $ for $i=1...r$
(see \eqref{eq:defdu}, the definition of $u_i$) using the following substitution for the generating function
\begin{equation*}
u=f_1(x^1+\tilde{x}^1)+g_1(x^1-\tilde{x}^1)+...+f_r(x^r+\tilde{x}^r)+g_r(x^r-\tilde{x}^r)
\end{equation*}
the condition \eqref{eq:s8} has the form
\begin{equation*}
-(U(x)-\Util(\tilde{x}))=f_1'(x^1+\tilde{x}^1)g_1'(x^1-\tilde{x}^1)+...+
f_r'(x^r+\tilde{x}^r)g_r'(x^r-\tilde{x}^r)
\end{equation*}

We are interested in functions for which the combinations such as $f_i'(x^i+\tilde{x}^i)g_i'(x^i-\tilde{x}^i)$ separate 
into sum
of two terms, one being only a function of $x^i$ and the other only of $\tilde{x}^i$.
In general case we still have \eqref{eq:s1} to \eqref{eq:s2} which give a nontrivial constraint.

\section*{Acknowledgments}

O.A. would like to thank Paul Windey for early discussions about this 
work.  This work was supported in part by
National Science Foundation grants PHY--0244261 and PHY--0554821.

% Put in bibliography style
% \bibliographystyle{utphys}
% \bibliographystyle{h-elsevier2}
% Put in my bibliography
% \bibliography{oabib}

\begin{thebibliography}{10}

\bibitem{Alvarez:2000bh}
O.~Alvarez, ``Target space duality. {I}: General theory,'' {\em Nucl. Phys.}
  {\bf B584} (2000) 659,
\href{http://www.arXiv.org/abs/hep-th/0003177}{{\tt hep-th/0003177}}.
%%CITATION = NUPHA,B584,659;%%.

\bibitem{Alvarez:2000bi}
O.~Alvarez, ``Target space duality. {II}: Applications,'' {\em Nucl. Phys.}
  {\bf B584} (2000) 682,
\href{http://www.arXiv.org/abs/hep-th/0003178}{{\tt hep-th/0003178}}.
%%CITATION = NUPHA,B584,682;%%.

\bibitem{Buscher:1988qj}
T.~H. Buscher, ``Path integral derivation of quantum duality in nonlinear sigma
  models,'' {\em Phys. Lett.} {\bf B201} (1988)
466.
%%CITATION = PHLTA,B201,466;%%.

\bibitem{Witten:1984ar}
E.~Witten, ``Nonabelian bosonization in two dimensions,'' {\em Commun. Math.
  Phys.} {\bf 92} (1984)
455--472.
%%CITATION = CMPHA,92,455;%%.

\bibitem{Kiritsis:1991zt}
E.~B. Kiritsis, ``Duality in gauged {WZW} models,'' {\em Mod. Phys. Lett.} {\bf
  A6} (1991)
2871--2880.
%%CITATION = MPLAE,A6,2871;%%.

\bibitem{Rocek:1992ps}
M.~Rocek and E.~Verlinde, ``Duality, quotients, and currents,'' {\em Nucl.
  Phys.} {\bf B373} (1992) 630--646,
\href{http://www.arXiv.org/abs/hep-th/9110053}{{\tt hep-th/9110053}}.
%%CITATION = NUPHA,B373,630;%%.

\bibitem{Gasperini:1993nz}
M.~Gasperini, R.~Ricci, and G.~Veneziano, ``A problem with nonabelian
  duality?,'' {\em Phys. Lett.} {\bf B319} (1993) 438--444,
\href{http://www.arXiv.org/abs/hep-th/9308112}{{\tt hep-th/9308112}}.
%%CITATION = PHLTA,B319,438;%%.

\bibitem{delaOssa:1993vc}
X.~C. de~la Ossa and F.~Quevedo, ``Duality symmetries from nonabelian
  isometries in string theory,'' {\em Nucl. Phys.} {\bf B403} (1993) 377--394,
\href{http://www.arXiv.org/abs/hep-th/9210021}{{\tt hep-th/9210021}}.
%%CITATION = NUPHA,B403,377;%%.

\bibitem{Giveon:1994ai}
A.~Giveon and M.~Rocek, ``On nonabelian duality,'' {\em Nucl. Phys.} {\bf B421}
  (1994) 173--190,
\href{http://www.arXiv.org/abs/hep-th/9308154}{{\tt hep-th/9308154}}.
%%CITATION = NUPHA,B421,173;%%.

\bibitem{Alvarez:1994qi}
E.~Alvarez, L.~Alvarez-Gaume, J.~L.~F. Barbon, and Y.~Lozano, ``Some global
  aspects of duality in string theory,'' {\em Nucl. Phys.} {\bf B415} (1994)
  71--100,
\href{http://www.arXiv.org/abs/hep-th/9309039}{{\tt hep-th/9309039}}.
%%CITATION = NUPHA,B415,71;%%.

\bibitem{Alvarez:1994zr}
E.~Alvarez, L.~Alvarez-Gaume, and Y.~Lozano, ``On nonabelian duality,'' {\em
  Nucl. Phys.} {\bf B424} (1994) 155--183,
\href{http://www.arXiv.org/abs/hep-th/9403155}{{\tt hep-th/9403155}}.
%%CITATION = NUPHA,B424,155;%%.

\bibitem{Sfetsos:1994vz}
K.~Sfetsos, ``Gauged {WZW} models and nonabelian duality,'' {\em Phys. Rev.}
  {\bf D50} (1994) 2784--2798,
\href{http://www.arXiv.org/abs/hep-th/9402031}{{\tt hep-th/9402031}}.
%%CITATION = PHRVA,D50,2784;%%.

\bibitem{Alvarez:2000pk}
O.~Alvarez, ``Target space pseudoduality between dual symmetric spaces,'' {\em
  Nucl. Phys.} {\bf B582} (2000) 139,
\href{http://www.arXiv.org/abs/hep-th/0004120}{{\tt hep-th/0004120}}.
%%CITATION = NUPHA,B582,139;%%.

\bibitem{Alvarez:2002mg}
O.~Alvarez, ``Pseudoduality in sigma models,'' {\em Nucl. Phys.} {\bf B638}
  (2002) 328--350,
\href{http://arXiv.org/abs/hep-th/0204011}{{\tt hep-th/0204011}}.
%%CITATION = HEP-TH 0204011;%%.

\end{thebibliography}

\providecommand{\href}[2]{#2}\begingroup\raggedright\endgroup

\end{document}